\documentclass[a4paper,10pt]{article}

\pdfoutput=1

\usepackage{cite,amsmath,amsfonts,amsthm,fullpage}
\usepackage{youngtab}

\usepackage{graphics}
\usepackage{miniplot}
\usepackage{subfigure}

\newcommand{\Pa}{\mathop\mathrm{P}\nolimits}

\newcommand{\DP}{\mathop\mathrm{DP}\nolimits}
\newcommand{\OP}{\mathop\mathrm{OP}\nolimits}

\newcommand{\ttr}{\texttt{tr}}

\theoremstyle{plain}

\newtheorem{Lemma}{Lemma}
\newtheorem{Proposition}{Proposition}
\newtheorem{Corollary}{Corollary}
\newtheorem{Remark}{Remark}
\newtheorem{Example}{Example}

\theoremstyle{remark}

\def\ttr{\mathrm {tr}}
\def\det{\mathrm {det}}

\def\diag{\mathrm {diag}}

\def\bp{\begin{Proposition}}
\def\ep{\end{Proposition}}
\def\bc{\begin{Corollary}}
\def\ec{\end{Corollary}}
\def\bl{\begin{Lemma}}
\def\el{\end{Lemma}}
\def\be{\begin{equation}}
\def\ee{\end{equation}}
\def\br{\begin{Remark}\rm\small}
\def\er{\end{Remark}}
\def\brs{\begin{remarks}.\\ \rm\
\begin{enumerate}}
\def\ers{\end{enumerate}\end{remarks}}
\def\bea{\begin{eqnarray}}
\def\eea{\end{eqnarray}}

\def\bx{\begin{Example}\rm\small}
\def\ex{\end{Example}}

\def\tr{\mathrm {tr}}
\def\det{\mathrm {det}}

\def\diag{\mathrm {diag}}

\def\&{&{\hskip -20pt}}

\newcount\YDcount\YDcount=0
\def\YDsize{10pt}

\def\YD#1{%
\ifnum#1=0
 \ifnum\YDcount=0 \ifx\varnothing\undefined\emptyset\else\varnothing\fi
 \else\vskip1.4pt\egroup\YDcount=0\fi
\else
 \ifnum\YDcount=0 \YDcount=1\vcenter\bgroup\vskip1pt
 \else\nointerlineskip\fi
 \vbox{\hrule\hbox{\vrule height\YDsize
 \loop\hskip\YDsize\vrule\ifnum\YDcount<#1\advance\YDcount1\repeat}\hrule
 \kern-0.4pt}\expandafter\YD
\fi}

\usepackage[usenames,dvipsnames]{color}
\usepackage{ulem}

\def\pb{\mathbf{p}}
\def\tb{\mathbf{t}}

\def\zb{\mathbf{z}}
\def\zbb{\bar{\mathbf{z}}}

\def\zb{\mathbf{z}}

\def\Zb{\mathbf{Z}}

\def\i{\in\Zb}

\begin{document}

\author{ A.Yu. Orlov\thanks{Institute of Oceanology, Nahimovskii Prospekt 36,
Moscow 117997, Russia; National Research University Higher School of Economics
email: orlovs@ocean.ru}}

\title{Coupling of different solvable ensembles of random matrices II. Series over fat partitions: matrix models and discrete ensembles}

\maketitle

\begin{abstract}
We consider series over Young diagrams of products of Schur functions $s_{\lambda\cup\lambda}$, marked with ``fat partitions'' $\lambda\cup\lambda$, which appear in matrix models associated with ensembles of symplectic and orthogonal matrices and quaternion Ginibre ensembles. We consider mixed matrix models that also contain complex Ginibre ensembles labeled by graphs with corner matrices and the three ensembles mentioned above. Cases are identified when a series of perturbations in coupling constants turn out to be tau functions of the DKP hierarchy introduced by the Kyoto school. This topic relates matrix models to random partitions - discrete symplectic ensemble and its modifications.
\end{abstract}

\bigskip
\medskip


\section{Introduction \label{Introduction}}

The theory of integrable systems, to the development of which Vladimir Zakharov made a huge contribution, was initially aimed at describing nonlinear wave processes. However, it soon became clear that the theory created was of great importance in a wide variety of areas of mathematics and physics. One of the remarkable applications is the description of a number of matrix models associated with ensembles of random matrices. Matrix models arose independently in several areas, for example, in problems of quantum chaos \cite{Wigner},\cite{GorkovEliashberg}, statistical physics \cite{ItzyksonZuber}, string theory \cite{BrezinKazakov},
problems of information transmission \cite{AlfaroAkeman}. 
 For an overview, see the books \cite{Mehta}, \cite{Forrester}.
We distinguish two special cases, which we will call "solvable models" and "exactly solvable models". The term "solvable models" was introduced by Vladimir Kazakov in \cite{Kazakov-SolvMM}. These are models in which the statistical sum - the integral over some matrix ensemble - can be calculated as an explicit series over partitions of known polynomials - the Schur functions \cite{Mac}. 
To study solvable matrix models the saddle point method can be applied \cite{KazakovSW},\cite{Kazakov-SolvMM}.

Exactly solvable models are models in which the matrix integrals are tau functions of some hierarchy of classical integrable equations\footnote{The concept of tau function was introduced in the works of the Kyoto school \cite{JM} and turned out to be very fruitful}.
It is expected that exactly solvable models will have a number of remarkable properties and simplifications, so the search for such models is no less important a task than the search for new families of integrable equations. 
Many cases of exactly solvable matrix models have been studied, for matrix models related to KP,
TL nad multicomponent hierarchies see

\cite{GMMMO},\cite{GMMMMO},\cite{ZKMMO},\cite{Kontsevich} 
\cite{KMMM},\cite{MMSemenoff},\cite{LSpiridonov},\cite{KKZabrodinWiegman},\cite{HO2003},
\cite{ZinZub},\cite{O-2004-New},\cite{Oshiota2004},\cite{HO-JP-A},\cite{HO-tmp-2007}, 
\cite{Harnad-2014},\cite{Chekhov-2014},\cite{AOV},\cite{O-round-dance},\cite{MirMorCharact} and
also see the textbook \cite{BHarnad}.
For matrix models associated with other integrable systems, see
\cite{ChauZabor},\cite{AMS},\cite{Leur},\cite{OrlovGinibre},\cite{OST-II}, 
\cite{LOcharacter},\cite{NO-LMP},\cite{NO2020},\cite{NO2020tmp},\cite{MMNO},
\cite{SolvIII}. 

This note is related to the previous work \cite{AOV2} and is devoted to the search for new solvable and exactly solvable matrix models.
Exactly solvable matrix ensembles considered in \cite{AOV2} were related to the KP hierarchy integrated
by Vladimir Zakharov in 1974 \cite{ZS}. Here we consider ensembles related to the DKP hierarchy
\cite{JM}, which are certain reductions \cite{LOdetPfaff} of the matrix KP equations considered in \cite{ZS}.

In this paper we consider {\it mixed} ensembles: a mixed ensemble includes the multimatrix complex Ginibre ensemble, which we denote by $\Omega_n$ (where $n$ is the number of pairs of Hermitian conjugate complex random matrices), and one of three ensembles: (i) the ensemble $\omega_1$ of symplectic matrices $Sp(N)$, (ii)  the quaternion Ginibre ensemble $\omega_3$ , (iii) the ensemble $\omega_2$ of orthogonal matrices $O(N)$.
We show that such mixed ensembles lead to an interesting object: sums of products of Schur functions over "fat" Young diagrams and also sums over partitions with even parts. We also highlight cases where such matrix ensembles are tau functions, i.e., related to integrable systems, namely the DKP hierarchy \cite{JM}.

In our previous papers \cite{NO2020}, \cite{NO2020tmp}, \cite{AOV} we introduce a family of solvable
matrix models associated with embedded graphs (or, equivalently, ribbon graphs). Specifically,
to each embedded graph with $n$ ribbon edges, $F$ faces, and $V$ vertices, we associate an
$n$-matrix model (associated with the complex Ginibre multi-matrix ensemble) whose
partition function depends on $F$ sets of coupling constants $\pb^{(i)}=\left(p_1^{(i)},p_2^{(i)},\dots \right)$, $i=1,\dots,F$, and $V$ sets of constant matrices ${\cal V}_i$, $i=1,\dots,V$, see 
(\ref{ZU-MM}) below. We call these constant matrices vertex monodromies. They play a central role in what follows.

The goal of both \cite{AOV2} and the present note is to enlarge the set of solvable 
and the set of exactly solvable matrix models by the additional avaraging over vertex monodronomies which are no more constants but belong to 
ensambles $\omega_1,\omega_2$ or $\omega_3$. Thus, we deal with a new matrix model based on mixed
ensembles which turn up to be still solvable under additional avaraging. The subject of the present 
paper are matrix models which are series over partitions of a special type: partitions
of form $\lambda \cup \lambda$ which we called in \cite{OST-I} fat partitions.
Sums over such partitions generalize the model of random partition introduced in \cite{BorodinStrahov}. I believe that such sums will find their applications apart of 
topics of matrix models and random partitions.

The first two sections serve as an introduction and also provide a review of some necessary results from \cite{OS-TMP},\cite{OST-I},\cite{OST-II},\cite{NO2020} and \cite{AOV}. The results of this work are presented in the Section \ref{mixed}.

\section{Random matrix ensembles and matrix models \label{RME,MM}}

In this section we recall some important facts that were used earlier in constructing matrix models related to the DKP hierarchy
\cite{OST-II}, \cite{LOcharacter}, \cite{OrlovGinibre}. These ensembles
are not mixed, but "mono-ensembles" (which are related to integrals over one matrix).

We remind that the Schur function $s_\lambda$ is the polynomial function defined by
the partition $\lambda=(\lambda_1,\lambda_2,\dots)$ as follows
\be\label{s(p)}
s_\lambda(\pb)=\det\left[ s_{\lambda_i-i+j}(\pb)\right]_{i,j\ge 0},\quad 
e^{\sum_{m>0}\frac 1m p_m z^m}=\sum_{m\ge 0}z^m s_{(m)}(\pb)
\ee
In the special case where $\pb$  is chosen in the following way:
\be
\pb=\pb(X)=(\tr X,\tr\left( X^2 \right),\tr\left( X^3 \right),\dots ),
\ee
(where $X$ is an $N\times N$ matrix),
we write $s_\lambda(\pb(X))=:s_\lambda(X)$, using bold and capital letters to make the difference.
The Schur function $s_\lambda(X)$ is also known as the character of the linear 
group $\mathbb{GL}_N$ evaluated at $X\in\mathbb{GL}_N$ in representation $\lambda$. It is the symmetric polynomial in the eigenvalues $x_1,\dots,x_N$ of $X$:
\be
s_\lambda(X)=\frac{\det\left[x_j^{\lambda_i-i+N}\right]_{i,j}}{\det\left[x_j^{-i+N}\right]_{i,j}} .
\ee
The important fact is that the Schur function $s_\lambda(X)$ is also a polynomial of degree $|\lambda|$ in the entries of matrix $X$,
see Appendix \ref{Partitions},
and therefore it can be integrated over matrix elements, see (\ref{Schur-mean-SE}),
(\ref{Schur-mean-OE})  and
 (\ref{Schur-mean-qGE}) below. 
Let us note that $s_\lambda(X)=0$ in the event that the length of the partition 
(i.e., the number of its non-zero parts) exceeds $N$.

 The notation $\lambda\cup\lambda$ is used for partitions of the form $(\lambda_1,\lambda_1,\lambda_2,\lambda_2,\dots)$. The following series of Schur functions
 \be\label{DKPhyp-tau}
 \tau^{\rm DKP}(\pb,N)=\sum_{\lambda\cup\lambda\atop \ell(\lambda\cup\lambda)\le N} 
 s_{\lambda\cup\lambda}(\pb)\prod_{(i,j)\in\lambda\cup\lambda} r(j-i)
 \ee
 where $r$ is any function of one variable will be of use. 
 The product is taken over all nodes with coordinates $(i,j)$ in the Young diagram $\lambda\cup\lambda$.
 The series (\ref{DKPhyp-tau}) belongs to the family of hypergeometric DKP tau functions studied in \cite{OST-I} -
 a simple but very useful class of tau functions of the DKP hierarchy. The DKP hierarchy (KP hierarchy on the root system D) was introduced and studied by the Kyoto school, see \cite{JM}. It can be viewed \cite{LOdetPfaff} as a two-component KP hierarchy (a special case of the hierarchy introduced by Zakharov and Shabat \cite{ZS}) after applying some (orthogonal) reduction. The series (\ref{DKPhyp-tau}) we call hypergeometric because for rational $r$ the product $\prod_{(i,j)\in\lambda} r(j-i)$ can be written as a rational expression of Pochhammer symbols (see \cite{OS-TMP}), and $s_\lambda(X),\,|\lambda|=n$ is an analogue of $x^n$ and in this sense generalizes the usual Gaussian hypergeometric functions.

The following relation is called the Cauchi-Littlewood identity:
\be\label{CL}
e^{\sum_{m>0}\frac1m p_m \tr\left(X^m \right)}=\sum_{\lambda} s_\lambda(X)s_\lambda(\pb).
\ee
The matrix ensemble, say $\omega$, is given by a space of matrices (it may be Hermitian, symmetric, skew-symmetric, quaterninion self-dual, anti-self-dual, complex, real, unitary, orthogonal, or symplectic matrices), and a given invariant probabilistic measure on this space.

Then the expectation value 
\be\label{aMM}
Z(\pb)=\langle e^{\sum_{m} \frac1m p_m\tr \left(M^m \right)} \rangle_\omega = 
\sum_\lambda \langle s_\lambda(M)\rangle_\omega s_\lambda(\pb)
\ee
is called the partition function of a matrix model, or just a matrix model related to the matrix ensemble $\omega$. It depends on the coupling constants $\pb$. The exponential
can be replaced by another spectral invariant function.

They are also multi-matrix models, where the ensembles can be either the same, or they can be
different and the last case is our focus.

We need the expectations of the Schur functions $\langle s_\lambda(M)\rangle_\omega$ in various 
ensembles, and there are a number of works where such expectation values were evaluated.
Let me make references only to the very last mentioning of this problem \cite{MirMorCharact}.

Let me write down the examples we need.

(i) Let $M\in\mathbb{S}p(N)$, $N=2k$. Then
\be\label{Schur-mean-SE}
\langle s_\lambda(M) \rangle_{\omega_1}=
\int_{\mathbb{S}p(N)} s_\lambda(M)d\nu_1(M)=\begin{cases} 1\quad \lambda\in\mathbb{FP},\quad\ell(\lambda)\le N\\ 0 \quad {\rm otherwise}
\end{cases}.
\ee
Here $d\nu_1(M)=d_*M$ is the Haar measure on $\mathbb{S}p(N)$ (see Appendix \ref{ME}), and $\mathbb{FP}$ denotes the set of all
fat partitions.)

(ii) Let $M$ be an $N\times N$ matrix with quaternion entries that belongs to the quaternion-real 
Ginibre ensemble $q{\mathbb{GIN}}(N)$ (sometimes it is denoted by GinSE \cite{Forrester}). Then, for integer $L$, we have \cite{ForrRaine}:
\be\label{Schur-mean-qGE}
\langle s_\lambda(M)\det M^L \rangle_{\omega_2}=
\int_{q\mathbb{GIN}(N)} s_\lambda(M)\det(M^L)d\nu_2(M)=\begin{cases} N^{-|\lambda|}(N+L)_\lambda\quad \lambda\in \mathbb{FP}\\ 0 \quad {\rm otherwise}
\end{cases}.
\ee

(iii) Let $M\in\mathbb{O}(N)$, where $N$ is equal either to $2k$ or to $2k-1$. Then
\be\label{Schur-mean-OE}
\langle s_\lambda(M) \rangle_{\omega_3}=\int_{\mathbb{O}(N)} s_\lambda(M)d\nu_3(M)=\begin{cases} 1\quad \lambda\in 2\mathbb{P},\quad \ell(\lambda)\le n \\ 0 \quad {\rm otherwise}
\end{cases}
\ee
Here $d\nu_3(M)=d_*M$ is the Haar measure on $\mathbb{O}(N)$ (see Appendix \ref{ME}), and $2\mathbb{P}$ denotes the set of all partitions with even parts.

It is suitable to introduce notations: 
\be\label{rho}
\langle s_\lambda(M) \rangle_{\omega_i}=:\rho_i(\lambda)
\ee
where $\rho_{\omega_2}$ also depends on $N$ and $L$.

Then, the matrix models related to ensembles symplectic matrices are
\be\label{symplMM}
\langle e^{\sum_{m>0} \frac 1m p_m\tr \left(M^m \right)}\rangle_{\omega_1}=
\sum_{\lambda\atop\ell(\lambda)\le N} s_{\lambda\cup\lambda}(\pb)
\ee
and
\be\label{orthMM}
\langle e^{\sum_{m>0} \frac 1m p_m\tr \left(M^m \right)}\rangle_{\omega_2}=
\sum_{\lambda\in 2\mathbb{P}\atop \ell(\lambda)\le n} s_\lambda(\pb)=
\sum_{\lambda\atop\lambda_1\le  n} s_{\lambda\cup\lambda}(-\pb),
\ee
which both are tau DKP functions; see \cite{OST-I}, \cite{LOcharacter}.
Tau functions solve differential and difference equations in the variables $\pb$ and
discrete variables $N$ and $L$; see \cite{KvdLbispec}.

Although the quaternion-real Ginibre ensemble is described in the literature, it is a more involved case, and to better understand the matrix integral, we present the details:

\paragraph{Quaternion-real Ginibre ensemble.}

Quaternion Ginibre ensemble was introduced in \cite{Ginibre}. Details can be found in 
\cite{Mehta}, Ch 15.2, and \cite{Forrester}, \cite{ForrRaine}.

The quaternion Ginibre ensemble is the ensemble of $N\times N$ matrices with real quaternionic enties independently destributed according to the Gaussian measure.
One can see it as a $2N\times 2N$  matrix with the block structure, each block being presented 
as a sum with real coefficients over $\sigma$ matrices:

\be
{\bf M}_{ij} = \left[\begin{array}{cc}a&b\\c&d\end{array}\right]_{ij}=
\sum_\alpha M_{ij}^{(\alpha)}\sigma_{\alpha} ,
\ee
\be
 \sigma_1= \left[\begin{array}{cc}i&0\\0&-i\end{array}\right],
\sigma_2= \left[\begin{array}{cc}0&1\\-1&0\end{array}\right], \sigma_3= \left[\begin{array}{cc}0&i\\i&0\end{array}\right], \sigma_0= \left[\begin{array}{cc}1&0\\0&1\end{array}\right].
\ee
The measure on the space of real quaternion matrices is 
 \be\label{quatr-measure}
d\nu(M)=C \left(\prod_{{k \le j}}^N d M_{kj}^{(0)}\prod_{\alpha=1}^3\prod_{{k \le j}}^N d M_{kj}^{(\alpha)}\right)\text{e}^{-\frac N2 \tr \left(M^2\right)}  ,
\ee
where the normalization constant $C$ is chosen to provide $\int d\nu(M)=1$.

In \cite{OrlovGinibre}, the following matrix model 
\be\label{qGE}
\tau(N\pb,N,L)=\int e^{\sum_{m>0}\frac Nm p_m\tr\left( M^m \right)} \det M^L d\nu(M)
\ee
(here $L=0,1,\dots$)
was considered;  it was shown that
 the integral (\ref{qGE}) is the DKP tau function introduced in \cite{JM}, where the variables $t_i=\frac Ni p_i$ and discrete $L$ play the role of the DKP higher times.

Taking into account that the eigenvalues of $M$ enter by complex conjugated pairs $(z_i,\bar{z}_i),\,i=1,\dots,N$ and using the expression for joint probability function (formula (15.2.10) in \cite{Mehta}),
this integral can be re-written to form the following integral over $n$ pairs of complex 
variables $z_i,\bar{z}_i$:
\be\label{tauGin}
\tau(N\pb,N)=\int_{\mathbb{C}^n} \Delta(\zb,\zbb)
\prod_{i=1}^n e^{-N|z_i|^2+\sum_{m>0} \frac Nm p_m\left(z_i^m + {\bar z}_i^m \right)}
|z_i-{\bar z}_i|^2|z_i|^{2L} d^2z_i ,
\ee
where
\be
\Delta(\zb,\zbb)=\prod_{i<j}|z_i-z_j|^2|z_i-\bar{z}_j|^2 .
\ee
Using the definition of Schur functions $s_\lambda$ and the Cauchy-Littlewood relation (see Appendix)
one can rewrite (\ref{tauGin}) as 
\be\label{hyptau}
\tau(N\pb,N)=\sum_{\lambda\atop\ell(\lambda)\le N} N^{-|\lambda|}(N+L)_\lambda s_{\lambda \cup \lambda}(N\pb) ,
\ee
where
\be\label{Nlambda}
(a)_\lambda:=\prod_{(i,j)\in\lambda} (a+j-i)
\ee
is the Pochhammer symbol related to the partition $\lambda=(\lambda_1,\dots,\lambda_N)$
and can be also written as $(a)_{\lambda_1}(a-1)_{\lambda_2}\cdots (a-N+1)_{\lambda_a}$ where 
$(a)_k=\frac{\Gamma(a+k)}{\Gamma(a)}$ is the ordinary Pochhammer symbol.
Such a series is an example of hypergeometric DKP tau functions considered in \cite{OST-I}.

\br 
The so-called orthogonal and symplectic ensembles (they got their name from the type of symmetry of the measure, see \cite{Mehta}) should not be confused with the ensemble of orthogonal matrices $\omega_3$ and the ensemble of symplectic matrices $\omega_1$. These are different matrix ensembles! In the first part of our paper \cite{AOV2} we consider the interaction of unitary, orthogonal and symplectic ensembles with the complex multi-matrix Ginibre ensemble. The mixed ensembles considered in \cite{AOV2} do not belong to DKP, but to the KP hierarchy and to the "large BKP" hierarchy introduced by Kac and van de Leur.
\er

\br
The series (\ref{tauGin}) for the matrix model (\ref{qGE}) may be compared with the series
\be\label{1MMSchur}
\tau^{\rm KP}(\pb,N)=\sum_{\lambda\atop\ell(\lambda)\le N} 
(N)_\lambda s_\lambda(\pb')s_\lambda(0,1,0,0,\dots)
\ee
obtained  in \cite{HO2003}
for the Kazakov-Brezin-Gross-Migdal \cite{BrezinKazakov},\cite{GrossMigdal} 
matrix model, where $M$ is the Hermitian matrix:
\be\label{1MM}
\int e^{\sum_{m>0}\frac 1m p_m\tr\left(M^m\right)} e^{-c\tr\left(M \right)^2} dM,\quad 
dM=\prod_{i\le j} d\Re M_{ij}\prod_{i < j} d\Im M_{ij}
\ee
In (\ref{1MMSchur}) $\pb=(p_1,p_2,p_3,\dots)$ and $\pb'=(p_1,p_2-\frac 14,p_3,\dots)$.
\er
\br\label{polynom}
The series in the right-hand side can be either divergent or convergent in dependence of
the choice of the infinite set of coupling constants $\pb=(p_1,p_2,\dots)$. In case we choose
\be\label{pol}
p_k=-\sum_{i=1}^{N'} x_i^k
\ee
with any choice of the complex parameters $x_i,\,i=1,\dots,N'$ and the number $N'$, then 
the right-hand side is a polynomial, namely, a polynomial of degree $NN'$ if we 
put $\deg p_k=k$.
\er

\section{Models of $n$ pairs of complex matrices with $2n$ sources \label{ComplexGinibre} \cite{NO2020}}

Let us recall, in short, the construction in \cite{NO2020}, \cite{AOV}.

 An embedded graph is a graph whose faces are homeomorphic to a disk.
Consider an embedded graph (also known as a ribbon or fat graph) $\Gamma$ with $F$ faces, $n$ edges, and $V$ vertices.
We number each edge with a pair of numbers $i$ and $-i$ ($i=1,\dots,n$), assign these numbers to the sides of the edge, it doesn’t matter to which side we assigned $i$ and to which $-i$, but it’s important for us to fix this.
We assign matrices $Z_a\in Mat_{N\times N}$ to the side $a$ ($a=\pm 1,\dots,\pm n$), and we impose $Z_a=Z^\dag_{-a}$. (The collection of the matrices $\{Z_a\},\,a=\pm 1,\dots,\pm n$
we will be used as the collection random matrices, see below.)
We number each face of $\Gamma$ and consider each border side of a face as positively 
oriented. Let's go around the edge of the face in the positive direction.
We number each corner of the face that follows the side $a$ by the same number.
 We assign an $N\times N$ matrix $M_a$ to a corner $a$. 
 Let us number the faces by $i=1,\dots,F$
 We introduce the monodromy 
of a face as a consequent product of corner matrices from the set $\{M_a,\,a=\pm 1,\dots,\pm n\}$ when we go around the border of this face  in the positive direction. 
In addition, we introduce the "monodromy dressed by random matrices" of a face as a successive product of pairs $Z_aM_a$ as we move along the boundary of the face.
Let us number each face as $i=1,\dots,F$, the monodromy of face $i$ and the face monodromy dressed by random matrices are denoted by ${\cal F}_i$ and ${\cal F}_i(X)$, respectively.
 Next, let us number the vertices of $\Gamma$ by $1,\dots,V$, and consider a vertex $i$.
Let us introduce the monodromy of the vertex $i$ as the consequent product of adjacent corner matrices when we go around it in a negatice (clockwise) direction.
Let us denote the monodromy of vertex $i$ by ${\cal{V}}_i$. 
Both monodromies are defined up to
a cyclic permutation.

Let us consider two the most simplest examples:

(1) $\Gamma=\Gamma_1$ is the segment: $F=1,n=1,V=2$. We have the pair of random matrices $Z_1$ and $Z_{-1}=Z_1^\dag$ attached to the sides of the single edge,
and two corner matrices at the ends, which are one-valent vertices and related "corners" are numbered by $1$ and $-1$. If we go around the boundary of the face in the positive 
direction we consequently meet $Z_1$, then $M_1$, then $Z_{-1}=Z_1^\dag$, then $M_{-1}$.
The face monodromy is ${\cal F}=M_1M_2$, the dressed face monodromy is ${\cal F}(X)=Z_1M_1Z_{-1}M_{-1}$. There are two vertex monodromies ${\cal V}_1=M_1$ and ${\cal V}_2=M_{-1}$.

(2) $\Gamma=\Gamma_2$ is now dual to the previous graph. It is an edge that starts and ends at a single vertex (this edge can be thought of as the equator on the globe).
$F=2, n=1, V=1$. We get ${\cal F}_1=M_1$ and ${\cal F}_2=M_{-1}$. Thus, ${\cal F}_1(X)=Z_1M_1$
and ${\cal F}_2(X)=Z_{-1} M_{-1}$. Also ${\cal V}=M_1M_{-1}$.

The simplest graph on a closed orientable surface of genus $g$ is the graph whose ribbon 
edges are $a$ and $b$ cycles, and whose only face is the fundamental domain at the $4g$ corners of which the corner matrices are placed. The reader, as an exercise, can write out the dressed monodromy ${\cal F}^{(g)}(X)$ of this face and the monodromy ${\cal V}^{(g)}$ of the single vertex of the related graph. We are most interested in the case $g=0$.

Thus, we have the set of complex matrices $Z_{\pm 1},\dots,Z_{\pm n}$ and a set of corner
matrices $M_{\pm 1},\dots,M_{\pm n}$ (in \cite{NO2020},\cite{NO2020tmp},\cite{AOV} we call them 
``source matrices'').

The collection of complex matrices $Z_{\pm i} \in\mathbb{GL}_N$, $i=1,\dots,n$ we denote by $X$.
The Ginibre ensemble of $n$ random complex matrices $X$ is defined by the following
measure
$$
dX=\prod_{a=1}^n d\mu(Z_a),\quad d\mu(Z_a)=C \prod_{i,j\le N}e^{-N |(Z_a)_{i,j}|^2} d^2 (Z_a)_{i,j}
$$
where $C$ is chosen in a way that $\int_{\mathbb{GL}_N\times\cdots \times\mathbb{GL}_N}dX=1$.
 We have the following wonderful relation \cite{NO2020tmp}:
\be\label{I}
\int_{\mathbb{GL}_N\times\cdots \times\mathbb{GL}_N}\prod_{a=1}^F s_{\lambda^a}\left({\cal F}_a(X)\right) dX=\delta_\lambda
\left(s_\lambda(N\pb_\infty)\right)^{-n}
\prod_{a=1}^V s_{\lambda^a}\left({\cal V}_a\right) ,
\ee
where $\delta_\lambda$ is equal to 1 in case $\lambda^1=\cdots = \lambda^F$ and where we
denote $\lambda^1$ by $\lambda$. And $\delta_\lambda=0$ otherwise. We recall that 
$N\pb_\infty=(N,0,0,\dots)$, then (see \cite{Mac})
\be
s_\lambda(N\pb_\infty)=N^{nd}\frac{d_\lambda}{d!},\quad d_\lambda:=
\frac{\prod_{i<j\le N}(\lambda_i-\lambda_j-i+j)}{\prod_{i=1}^N(\lambda_i-i+N)!},\quad 
d:=|\lambda|
\ee
(where $d_\lambda$ is the dimension of the representation $\lambda$ of the symmetric group $S_d$).

\br
Actually we also have
\be\label{I'}
\int_{\mathbb{GL}_N\times\cdots \times\mathbb{GL}_N}\prod_{a=1}^V s_{\lambda^a}\left({\cal V}_a(X)\right) dX=\delta_\lambda
\left(s_\lambda(N\pb_\infty)\right)^{-n}
\prod_{a=1}^F s_{\lambda^a}\left({\cal F}_a\right) ,
\ee
where ${\cal V}_a(X)$ denote the dressed face monodromies of the dual graph and now ${\cal F}_a$
are vertex monodromies of the dual graph.
\er

\paragraph{Matrix models.}
Now we consider the matrix model related to an embedded graph $\Gamma$ as
$$
{\rm Z}_N(N\tb; {\cal M} |\Gamma)=\int_{\mathbb{GL}_N\times\cdots \times\mathbb{GL}_N} \prod_{i=1}^F e^{\sum_{m>0}\frac Nm p^{(i)}_m \tr \left({\cal F}_i(X)\right)^m}dX=
$$
\be\label{ZU-MM}
=\sum_{\lambda\atop \ell(\lambda)\le N} 
\left(s_\lambda(N\pb_\infty)\right)^{-n}
\prod_{i=1}^F s_\lambda(N\pb^i)
\prod_{i=1}^V s_\lambda({\cal V}_i)
\ee
where $\tb$ denotes the collection of coupling constants $\pb^i,\,i=1,\dots,F$ and
${\cal M}$ denotes the collection of the source matrices $M_{\pm i},\,i=1,\dots,n$.
This model has $F$ sets of coupling constants $\pb^i=(p^{(i)}_1,p^{(i)}_2,\dots)$ and 
the set of $V$ independent combinations ${\cal V}_i$ constructed from the source matrices
(or, the same, from the corner matrices of the graph $\Gamma$). 
Note that the right-hand side is built only on Schur functions
\footnote{We pointed out the same phenomenon in the formula for Milne's hypergeometric functions, see \cite{OS-TMP}.}.

A beautiful equality (\ref{ZU-MM}) was proved in \cite{NO2020} using the combinatorial-geometric definition of Hurwitz numbers by gluing a covering of polygons and Cauchy-Littlewood (\ref{CLit}) and characteristic map \ref{charmap} formulas.
Note that if we assign degree one to each Schur function, then the total degree of
${\rm Z}_N(N\tb; {\cal M} |\Gamma)$ will be the Euler characteristic of the surface at which the graph of $\Gamma$ is embedded.

\br 
Sums over partitions of Schur function products, similar to the right-hand side in
(\ref{ZU-MM}), appeared in \cite{AMMN-2011}, \cite{AMMN-2014} without connection to matrix models (and without connection to embedded graphs) as a generating
function of Hurwitz numbers. 

\er

\br 
Relations between matrix integrals and generating functions for special combinations of Hurwitz numbers were presented earlier in the
\cite{Kostov}, which is closest to our work - it did not take into account the source matrices and the vertex monodromies. The formula (\ref{ZU-MM}) is a more powerful result obtained in a different approach taking into account the vertex monodromies,
which are additional free parameters of the model and allow generating not special combinations
of Hurwitz numbers, but the Hurwitz numbers themselves. There are a number of other works devoted to certain matrix integrals and special combinations of Hurwitz numbers:
\cite{Alexandrov},
as well as works on combinations of Hurwitz numbers and matrix integrals related to the KP equations in
\cite{Harnad-2014},
\cite{Chekhov-2014},
\cite{NO-2014}. For the Kontsevich matrix model as an exactly solvable matrix BKP model associated with special combinations of Hurwitz spin numbers, see \cite{MMNO}.
\er

\br \label{rectangular} As one can see from the expression for the face monodromies,
to get multi-matrix models with rectangular complex matrices, it is enough to choose
a corner matrix, say $M_i$, in form $$M_i={\rm diag}(1,1,\dots,1,0,\dots,0),$$ 
Since the random matrix $Z_i$ enters the integrand only as the product $Z_iM_i$, it can be replaced by a rectangular one.
\er

\br\label{Important}
Notice that each corner matrix enters only one of the vertex monodromy as a factor in the product of corner matrices around a vertex.
However the right hand side of (\ref{ZU-MM}) depends only on the set of vertex monodromies.
It means that we
have freedom to choose corner matrices keeping only their products around vertices and in addition  up to 
the permutation of vertex monodromies.
Let us call it the gauge freedom of the model.
As a result for a given numbers $n$, $F$ and $V$ and for a given set of ${\cal V}$
we have a subfamily of solvable models related 1) to the choice of the graph and 2) to
the choice of the gauge freedom. Each member of the subfamily of these matrix models has
same partition function given by (\ref{ZU-MM}).
\er

Thus, we can say this: the final answer for a family of models is given by a graph whose edges correspond to random complex matrices, and whose vertices (up to the permutations) correspond to constant given matrices (vertex monodromies). And the factorization of each vertex monodromy into each corner matrices and also the permutations of the monodormies in the given set ${\cal V}:={\cal V}_1,\dots,{\cal V}_V$ is our freedom to choose an integral in this family. Each family consists of solvable matrix integrals. 
 Thus, we can denote the integrals
as ${\rm Z}_N(N\tb; {\cal V} |F,n,V)$, however, if we want to specify the realization, we will write it as ${\rm Z}_N(N\tb; {\cal M} |\Gamma)$

\br
Both the left and the right sides of (\ref{ZU-MM}) can be either convergent or divergent depending
on the range of values of the parameters $\pb^{(i)}$. For many combinatorial problems,
convergence is not important: only the recurrence relations are important.
\er

Examples of multi-matrix models associated with various graphs (with pictures) can be found in \cite{AOV}.

However, one needs further restrictions to have exactly solvable families. 
The selection of exactly solvable families is described in \cite{AOV2}. In particular, the graph $\Gamma$ must be drawn on the Riemannian sphere. Any other closed Riemann surface does not give exactly solvable families.

\subsection{Selection of exactly solvable families among series (\ref{ZU-MM}\label{exact}).
} This problem was considered in \cite{AOV} and \cite{AOV2}.

Below we use formulas (\ref{s(p(a))}) and (\ref{s(J_l)}) of Appendix \ref{Partitions} and
also the formula 
\be\label{KPtau}
\sum_\lambda \left(s_\lambda(\pb_\infty)\right)^{-n}s_\lambda(X)s_\lambda(\pb)\prod_{i=1}^n s_\lambda(\pb(a_i))=
\sum_\lambda 
s_\lambda(\pb)s_\lambda(X)\prod_{i=1}^n (a_i)_\lambda=\tau^{2KP}(\pb,X)
\ee 
which belongs to a special family of KP tau functions, which we call hypergeometric KP tau functions, see \cite{OS-TMP}.

Thus, we compare the left side (\ref{KPtau}) with the right side (\ref{ZU-MM}).
We can choose some of $\pb^{(i)}$ as $\pb(a_i)$ and some of ${\cal V}_i$ as $J_{l_i}$ (keeping in mind that $s_\lambda(J_{l_i})=s_\lambda(\pb(l_i))$), since they are free parameters in 
(\ref{ZU-MM}).
We find the following cases:
\begin{itemize}
\item $F-n+V=2$ (For a certain modification of the integral on the left-hand side (\ref{ZU-MM}), which is beyond the scope of this topic, one can use
$F-n+V=1$)
\item   some of the sets $\pb^i$ are equal to $\pb_\infty=(1,0,0,\dots)$
\item some of the sets $\pb^i$ are equal to $\pb(a)=(a,a,\dots)$.
In this case \cite{Mac},
\be\label{p(a)}
s_\lambda(\pb(a))=s_\lambda(\pb_\infty)\prod_{(i,j)\in\lambda}(a+j-i)
\ee
\item the spectrum of some of ${\cal V}_i$  consists of $l_i\le N$ unities 
and $N-l_i$ zeros. We denote such matrices by $J_{l_i}$. In case $l_i=N$ it is the identity matrix
$I_N$.
Let us note that $s_\lambda(J_{l_i})$ is equal to zero in if
the length of $\lambda$ exceeds $N-l_i$.
\end{itemize}

More comments on such specification:

The case $l_i<N$ means that we are actually dealing with {\it rectangular random matrices}, since the face monodromy contains a product of corner (source) matrices, which must also be
singular. Then some products $Z_iM_i$ included in the face monodromy are singular and $M_i$ can be replaced by rectangular matrices under the trace sign.

The specification of $\pb^i$ means that
\be\label{p-infty}
{\rm for}\,\, \pb^i=\pb_\infty\,\, {\rm we \,have}\,\,
e^{\sum_{m>0}\frac Nm p^{(i)}_m \tr \left({\cal F}_i(X)\right)^m}=
e^{ N\tr {\cal F}_i(X)}\,,
\ee
\be\label{p(a)'}
\qquad\quad{\rm for}\,\, \pb^i=\pb(a_i)\,\, {\rm we \,have}\,\,
e^{\sum_{m>0}\frac Nm p^{(i)}_m \tr \left({\cal F}_i(X)\right)^m}=
\det\left(1-{\cal F}_i \right)^{-N a_i}\,,
\ee
where $a_i$ is a chosen parameter.

\bp\label{ex-solv}
Under any choice of restrictions above the integral (\ref{ZU-MM}) is the hypergeometric KP tau function, described by formula (3.12.6) in \cite{OS-TMP}.
\ep

The proof follows from the relations $\frac{s_\lambda(\pb(a))}{s_\lambda(\pb_\infty)}=(a)_\lambda$, see \cite{Mac}, and $\frac{s_\lambda(J_{m_i})}{s_\lambda(N\pb_\infty)}=N^{-n|\lambda|}(m_i)_\lambda$, which both have the form of the so-called content product defining the hypergeometric tau functions (an alternative way of defining these functions can be found in \cite{KMMM}).

These restrictions are sufficient to obtain exactly solvable partition functions given by (\ref{ZU-MM}). I have no evidence that these restrictions are necessary.

\br
Note that the simplest case for analyzing the series on the right-hand side of (\ref{ZU-MM}) is the case when all the listed constraints are satisfied simultaneously and at the same time
$ F-n+V = 0 $, which corresponds to the graphs drawn on the torus. In this case, the answer is a sum over partitions, each term of which is decomposed into functions of the shifted parts of the partitions $\lambda_i-i$,
 see \cite{OS-TMP}.
\er

\section{Mixed ensembles and series over fat partition \label{mixed}}

Mixed ensembles are ensembles of complex matrices and matrices from ensembles 
(i), (ii) and (iii). The partition function of the matrix model is obtained by the avaraging 
the partition function of the matrix models (\ref{ZU-MM})
over ${\cal V}_i\in \omega_{j_i},\, i=1,\dots,v\le V$, so that the resulting partition function
depends only on the set $\tb$ and on $\bar{\cal V}={\cal V}_{V-v},\dots,{\cal V}_V$, then we get:
$$
S(N\tb; \bar{\cal V} |F,n,V):=\int_{\mathbb{GL}_N\times\cdots \times\mathbb{GL}_N}\,
\langle \prod_{i=1}^F e^{\sum_{m>0}\frac Nm p^{(i)}_m \tr \left({\cal F}_i(X)\right)^m}\rangle_{\omega_{j_1},\dots,\omega_{j_v}}dX=
$$
\be\label{ZU-MM''}
=\sum_{\lambda\atop \ell(\lambda)\le N} 
\left(s_\lambda(N\pb_\infty)\right)^{-n}
\prod_{i=1}^F s_\lambda(N\pb^i)
\prod_{i=1}^{v} \langle s_\lambda({\cal V}_i) \rangle_{\omega_{j_i}}
\prod_{i=V-v}^{V} s_\lambda({\cal V}_i) ,
\ee
where each $\langle \rangle_{\omega_{j_i}}$ denotes taking the mathematical expectation in the ensemble $\omega_{j_i}$, each of $j_i$ is selected from the set $1,2,3$. Here $v$ is the number of additional averaging of the statistical sum (\ref{ZU-MM}).

Now, as can be seen from (\ref{Schur-mean-SE}) and (\ref{Schur-mean-qGE})
when averaging over the $v_2$ quaternionic and over the $v_1=v-v_2$ sympeltic ensembles,
the sum (\ref{ZU-MM}) gives
\be\label{ZU-MM''qGE}
\langle {\rm Z}_N(N\tb,{\cal V}|F,n,V) \rangle_{\omega_2^{\times v_2},\omega_1^{\times v_1}}=
\sum_{\lambda\cup\lambda \atop \ell(\lambda\cup\lambda)\le N} 
\left(s_{\lambda\cup\lambda}(N\pb_\infty)\right)^{-n}
\prod_{i=1}^F s_{\lambda\cup\lambda}(N\pb^i)
\prod_{i=1}^{v_2} \rho_{2}(\lambda)
\prod_{i=V-v}^{V} s_{\lambda\cup\lambda}({\cal V}_i) ,
\ee
where $v_2$ is the number of quaternionic matrices among vertex monodromies we avarage.
In case we avarage only over orthogonal matrices we get
$$
\langle {\rm Z}_N(N\tb,{\cal V}|F,n,V) \rangle_{\omega_3^{\times v_3}}=
\sum_{\lambda\in 2\mathbb{P} \atop \ell(\lambda)\le N} 
\left(s_{\lambda}(N\pb_\infty)\right)^{-n}
\prod_{i=1}^F s_{\lambda}(N\pb^i)
\prod_{i=V-v}^{V} s_{\lambda}({\cal V}_i) ,
$$
\be\label{ZU-MM''-orth}
=\sum_{\lambda\cup\lambda \atop \lambda_1\le N} 
\left(s_{\lambda\cup\lambda}(N\pb_\infty)\right)^{-n}
\prod_{i=1}^F s_{\lambda\cup\lambda}(-N\pb^i)
\prod_{i=V-v_3}^{V} s_{\lambda\cup\lambda}({\cal V}_i)
\ee
Here we use the fact that a partition with even parts is the conjugate to a fat partition
and
that $s_\lambda(\pb)=(-1)^{|\lambda|}s_{\lambda^t}(-\pb)$, where $\lambda^t$ is the conjugate of $\lambda$, see \cite{Mac}, $|\lambda|$ is the weight of $\lambda$ (namely, the weight is the sum of parts of $\lambda$), which in our case is an even number.

In the case of averaging over orthogonal matrices, as well as over symplectic and quaternion matrices, the summation range is limited by partitions $\lambda$, which are simultaneously both thick and even partitions.

At last one can study the right hand side of (\ref{ZU-MM''qGE}) in the context of discrete ensembles and random partitions, see \cite{BorodinStrahov}.

\paragraph{Selection of exactly solvable cases.}

The selection of exactly solvable cases leads not to the tau function of the KP hierarchy, but to the tau function of the DKP hierarchy.

\bp\label{ex-solv-mixed}
Suppose $v=1$.
If the integral (\ref{ZU-MM''}) is subject to the restrictions set out in Section \ref{exact}, then the series (\ref{ZU-MM''}) is the hypergeometric DKP tau function, given by formula (138) of the paper \cite{OST-I}.
\ep
The proof is the same as the proof of Proposition \ref{ex-solv} with the additional remark that $\rho_2(\lambda)$ defined by (\ref{rho}) and (\ref{Schur-mean-qGE}) which enter (\ref{ZU-MM''qGE})
also has the form of content product. Details about hypergeometric DKP tau functions see in \cite{OST-I}.

For a graph with $F$ faces $n$ edges and $V$ vertices we have $n$ different corners and one can place the matrix ${\cal V}$ we avarage to any of them. We will give a visual example in 
Section \ref{equalities}.

As in the case of Section \ref{exact} 
the restrictions we refer to are sufficient to obtain exactly solvable partition functions given by (\ref{ZU-MM''}). However I have no evidence that these restrictions are necessary.

\subsection{Examples of exactly solvable matrix models with mixed ensembles}

Here we will look at simple cases to understand how our construction works.

(a) Graph is a segment. In this case $F=1, n=1, V=2$.

Let $Z$ be a complex matrix and $Z^\dag$ it's Hermitian conjugate, then let $M$ is a matrix
that belongs to an ensemble $\omega_i$ from Section \ref{RME,MM} and $J_l$ is a matrix that has $n$ non-zero eignevalues 
equal to $1$ each, so for $l=N$, it is the identity matrix $I_N$.
Consider
the integral 
\be\label{gamma1}
S_i(N\pb;J_l|\Gamma_1)=\langle {\rm Z}_N(N\pb;{\cal V}|\Gamma_1)\rangle_{\omega_i}
=\int_{\omega_i}\int_{\mathbb{GL}_{N}} e^{\sum_{m>0} \frac Nm p_m\tr\left((Z^\dag M ZJ_l)^m\right)} d\mu(Z)d\nu_i(M) ,
\ee
where the notation $\Gamma_1$ is the graph, which is the segment drawn on the sphere (one face, one ribbon edge, two vertices); as it was explained in Section \ref{ComplexGinibre} the graph defines 
the integrand. 
Let us check it:
For such graph we have two corner matrices: $M$ and $J_l$.
We get the monodromy of the face, which is ${\cal F}=M J_l$ and the dressed monodromy of the face which is ${\cal F}(X)=Z^\dag M Z J_l$ (we used $X$ to denote the whole collection of random matrices 
assigned to the ribbon edges of a graph. In the present example $X$ consists of $Z$ and $Z^\dag$).
And thus, we are convinced that $\Gamma_1$ is indeed our graph which describes the integrand
and $\pb=(p_1,p_2,\dots)$ is the set of coupling constants assigned to the (single) face.
We have two vertex monodromies $M$ and $J_l$. Then before the avaraging over vertex monodromy $M$ 
we have the integral over complex matrix $Z$, according to (\ref{ZU-MM}) it is
$$
{\rm Z}_N(N\pb;M,J_l|\Gamma)=\sum_{\lambda\atop\ell(\lambda)\le N} \frac{s_\lambda(N\pb)s_\lambda(M)s_\lambda(J_l)}{s_\lambda(N\pb_\infty)}=
\sum_{\lambda\atop\ell(\lambda)\le N} (l)_\lambda s_\lambda(N\pb)s_\lambda(M) N^{-|\lambda|}
$$
which satisfies the restriction in Section \ref{exact} because $J_l$ is chosen to have $l$ eigenvalues equal to $1$ and rest ones are $0$. Then $\frac{s_\lambda(J_l)}{s_\lambda(N\pb)}=
N^{-|\lambda|}(l)_\lambda$ is the content product and ${\rm Z}_N(N\pb;M,J_l|\Gamma)$ is the KP 
hypergeometric tau function \cite{OS-TMP}. Now the avaraging of it over, say symplectic matrices 
$M\in\mathbb{S}p(2k)$ gives
\be\label{gamma1,1}
S_1(2k\pb|\Gamma_1)=
\sum_{\lambda\atop\ell(\lambda)\le 2k} (l)_{\lambda}\rho_1(\lambda) s_{\lambda}(2k\pb)(2k)^{-|\lambda|}=\sum_{\lambda\cup\lambda\atop\ell(\lambda)\le k}
(l)_{\lambda\cup\lambda} s_{\lambda\cup\lambda}(2k\pb)(2k)^{-|\lambda|}
\ee

In case $M\in q{\mathbb{GIN}}(2k)$ after evaluation we get
\be\label{gamma1,2}
S_2(2k\pb|\Gamma_1)=\sum_{\lambda\atop \ell(\lambda)\le k} (k)_{\lambda\cup\lambda} (l)_{\lambda\cup\lambda}(2k)_{\lambda\cup\lambda} s_{\lambda\cup\lambda}(N\pb) N^{-2|\lambda|}
\ee
The right hand sides of (\ref{gamma1,1}),(\ref{gamma1,2}),(\ref{gamma1,3})
are examples of the DKP tau function.

The similar answer we obtain for $M\in \mathbb{O}(N)$, say $N=2k$,
\be\label{gamma1,3}
S_3(2k\pb|\Gamma_1)=\sum_{\lambda \atop\lambda_1\le n} (-l)_{\lambda\cup\lambda} s_{\lambda\cup\lambda}(-2k\pb)
N^{-2|\lambda|}
\ee
where we took into account the relation 
$s_\lambda(\pb(l))=(-1)^{|\lambda|}s_{\lambda^t}(-\pb(l))=(-l)_{\lambda^t}$ and that $|\lambda|$ is an even number.

Another example with the same graph:
in (\ref{ZU-MM''}) instead of the constant corner matrix $J_l$ in the integral we take the random matrix $M_1\in\mathbb{S}p(2k)$, and $M\in\mathbb{S}p(2k)$. We obtain:
\be\label{gamma1full}
\tilde{S}_i(N\pb|\Gamma_1)=\int_{\mathbb{S}p(2k)}\int_{\mathbb{S}p(2k)}\int_{\mathbb{GL}_{N}}
 e^{\sum_{m>0} \frac Nm p_m\tr\left((Z^\dag M Z M_2)^m\right)} d\mu(Z)d\nu_1(M)d\nu_1(M_1)
\ee
$$
=\sum_{\lambda\cup\lambda\atop \ell(\lambda)\le k} 
\frac{s_\mu(2k\pb) }{s_\mu(2k\pb_\infty)}
$$
In our terminology this is solvable but not exactly solvable case.

(b) The graph is an equator with a marked point drawn on a sphere, so $F=2, n=1, V=1$.

We have
\be\label{gamma2}
S_i(\pb,N|\Gamma_2) =\int_{\omega_i}
\int_{\mathbb{GL}_{N}} 
e^{N\tr Z^\dag +\sum_{m>0}\frac Nm p_m \tr\left((ZM)^m\right)}
d\mu(Z)d\nu_i(M)
\ee
This integral can be viewed as a modification of the intgeral of Theorem 4 in \cite{O-2004-New}.
There are two faces with monodromies ${\cal F}_1=I_N$ and ${\cal F}_2=M$. The dressed 
monodromies are $Z^\dag$ and $ZM$. We have two sets of coupling constants related
to these faces. The vertex monodromy is ${\cal V}=M$. There we obtain
\be
S_i(N\pb^{(1)},N\pb^{(2)}|\Gamma_2)=\sum_{\lambda\atop \ell(\lambda)\le N} \frac{s_\lambda(N\pb^{(1)})s_\lambda(N\pb^{(2)})}{s_\lambda(N\pb_\infty)}\rho_i(\lambda)
\ee
In particular
\be
S_1(2k\pb^{(1)},2k\pb^{(2)}|\Gamma_2)=\sum_{\lambda\cup\lambda\atop \ell(\lambda)\le k} \frac{s_{\lambda\cup\lambda}(2k\pb^{(1)})s_{\lambda\cup\lambda}(2k\pb^{(2)})}{s_{\lambda\cup\lambda}(2k\pb_\infty)}
\ee
which is exactly solvable  in case 
$\pb^{(2)}=\pb(a)$. Then it is equal to
$$
\sum_{\lambda\cup\lambda\atop \ell(\lambda)\le k} (a)_{\lambda\cup\lambda}s_{\lambda\cup\lambda}(2k\pb^{(1)})(2k)^{-2|\lambda|}
$$
and in case $\pb^{(2)}=\pb_\infty$, then
\be\label{borodin}
\sum_{\lambda\cup\lambda\atop \ell(\lambda)\le k} (s_{\lambda\cup\lambda}(2k\pb^{(1)})
\ee
which are hypergeometric DKP tau function (equal to (\ref{gamma1,1}) in case $a=l$).

Examples (a) and (b) are the simplest.

(c) Graph $\Gamma_3$ is a segment with $p-1$ marked points - together with the ends of the segment 
they yield $V=p+1$ vertices; $F=1, n=p $.

It is the modified integral considered in \cite{Chekhov-2014}.  Namely, consider
\be\label{gamma3}
S_i(N\pb;J_{l_1},\dots,J_{l_p}|\Gamma_3)=\int_{\mathbb{GL}_N^{\times p}}\int_{\omega_i}
e^{\sum_{m>0} \frac Nm p_m
\left(Z_1\cdots Z_{p-1}Z_p M Z^\dag_pJ_{l_{p}}\cdots Z^\dag_1J_{l_1}\right)^m  }d\nu_i(M)\prod_{i=1}^p d\mu(Z_i)
\ee
Here $J_{l_i}=\diag(1,1,\dots,1,0,0,\dots 0)$, ${\rm rank}I_{l_i}=l_i$.
The face monodromy is ${\cal F}=I_N\cdots I_N M J_{l_{p}}\cdots J_{l_1}$ the dressed
monodromy is
${\cal F}(X)=Z_1\cdots Z_{p-1}Z_p M Z^\dag_pJ_{l_{p}}\cdots Z^\dag_1J_{l_1}$. 
The vertex monodromies,
 are ${\cal V}_1=J_{l_1},\dots,{\cal V}_p=J_{l_p}$ and ${\cal V}_{p+1}=M$.
If $M\in \mathbb{S}p(2k)$ we obtain after avaraging in ensemble $\omega_2$ :
\be\label{gamma3,1}
S_1(2k\pb;J_{l_1},\dots,J_{l_p}|\Gamma_3)=\sum_{\lambda\atop\ell(\lambda)\le N} 
s_{\lambda\cup\lambda}(N\pb) N^{-2p|\lambda|}\prod_{i=1}^p
(l_i)_{\lambda\cup\lambda}
\ee
which is DKP tau function.
We get similar formulas for ensembles $\omega_1$ and $\omega_3$. 

Let us compare it with the ensemble with the avaraging over all corner matrices
$$
S_{\rm full}(N\pb|\Gamma_3)=
\int_{\mathbb{GL}_N^{\times (p+1)}}\int_{\mathbb{S}p(2k)^{\times (p+1)}}
e^{\sum_{m>0}\frac Nm p_m
\left( (Z_1M_1)\cdots (Z_p M_p) (Z^\dag_p\cdots Z^\dag_1)\right)^m  }
\prod_{i=1}^{p+1} d\nu_1(M_i)\prod_{i=1}^{p} d\mu(Z_i)
$$
\be\label{*}
=\sum_{\lambda\cup\lambda\atop\ell(\lambda\cup\lambda)\le N} \left(s_{\lambda\cup\lambda}(N\pb_\infty)\right)^{-p}
s_{\lambda\cup\lambda}(N\pb)
\ee
This is not a tau function, though it is example of solvable matrix model.

\subsection{Equalities between partition functions of different matrix models\label{equalities}}

For a graph with $F$ faces, $n$ edges, and $V$ vertices, we have $n$ different corners, and we can place the average matrix ${\cal V}$ in any of them. However, this does not necessarily mean that we get different matrix models, since some of these integrals may turn out to be the same expressions after a change of notation. The same can be said about the choice of a face whose set of coupling constants will be the set of higher times of the DKP tau function. However, the same tau function can be {\it realized} as the partition function of matrix models that differ from each other: we have a set of models for the same DKP tau function.

How to write out different matrix models with the same statistical sum is clear from the construction of such models described above. However, it is useful to give an example. 

For simplicity, we will compare only ensembles with additional averaging over the group $\mathbb{S}p(N)$ ($N=2k$), and we will not compare integrals over the group $\mathbb{S}p(N)$ with averaging over GinSE.
Consider the case $F=2,n=2,V=2$. 
On the sphere with these data, there exist two different embedded graphs which the reader can easily keep in mind. 
Picking up exactly solvable cases according to the Proposition \ref{ex-solv-mixed}, and taking into account that for $\pb=\pb(a)$ we have
$$
e^{N\sum_{m} \frac{a}{m}\tr ({\cal F})^m}=\sum_\lambda s_\lambda(N\pb(a))s_\lambda( {\cal F}) =\det\left(1-{\cal F}  \right)^{-Na},
$$
and that $\frac{s_\lambda(J_l)}{s_\lambda(N\pb_\infty)}=(l)_\lambda N^{-|\lambda|}$ , we get:
\be\label{1}
\int e^{N\sum_{m>0} \frac{p_m}{m}\tr (Z_1M Z_2 )^m} \det\left(1-Z_1^\dag J_l Z_2^\dag\right)^{-Na}
d\mu(Z_1)d\mu(Z_2)d\nu(M)=
\ee
\be\label{2}
\int e^{N\sum_{m>0} \frac{p_m}{m}\tr (Z_1J_l Z_2 )^m} \det\left(1-Z_1^\dag M Z_2^\dag\right)^{-Na}
d\mu(Z_1)d\mu(Z_2)d\nu(M)
\ee
\be\label{3}
\int e^{N\sum_{m>0} \frac{p_m}{m}\tr (Z_1 Z_2 M )^m} \det\left(1-Z_1^\dag J_l Z_2^\dag\right)^{-Na}
d\mu(Z_1)d\mu(Z_2)d\nu(M)=
\ee
\be\label{4}
\int e^{N\sum_{m>0} \frac{p_m}{m}\tr (Z_1 Z_2 J_l)^m} \det\left(1-Z_1^\dag M Z_2^\dag\right)^{-Na}
d\mu(Z_1)d\mu(Z_2)d\nu(M)
\ee
\be\label{5}
=\int e^{N\sum_{m>0} \frac{p_m}{m}\tr (Z_1M Z_1^\dag Z_2)^m} \det\left(1- J_l Z_2^\dag\right)^{-Na}
d\mu(Z_1)d\mu(Z_2)d\nu(M)
\ee
\be\label{6}
=\int e^{N\sum_{m>0} \frac{p_m}{m}\tr (Z_1M Z_1^\dag J_l Z_2)^m} \det\left(1-  Z_2^\dag\right)^{-Na}
d\mu(Z_1)d\mu(Z_2)d\nu(M)
\ee
\be\label{7}
=\int e^{N\sum_{m>0} \frac{p_m}{m}\tr (Z_1M Z_1^\dag  Z_2 J_l)^m} \det\left(1-  Z_2^\dag\right)^{-Na}
d\mu(Z_1)d\mu(Z_2)d\nu(M)
\ee
\be\label{8}
=\int e^{N\sum_{m>0} \frac{p_m}{m}\tr (Z_2^\dag )^m} \det\left(1-   Z_1M Z_1^\dag J_l Z_2\right)^{-Na}
d\mu(Z_1)d\mu(Z_2)d\nu(M)
\ee
\be\label{9}
=\int e^{N\sum_{m>0} \frac{p_m}{m}\tr (Z_2^\dag)^m} \det\left(1-   Z_1M Z_1^\dag  Z_2 J_l \right)^{-Na}
d\mu(Z_1)d\mu(Z_2)d\nu(M)
\ee
\be\label{10}
=\int e^{N\sum_{m>0} \frac{p_m}{m}\tr (Z_1J_l Z_2 )^m} \det\left(1- M Z_2^\dag\right)^{-Na}
d\mu(Z_1)d\mu(Z_2)d\nu(M)
\ee
\be\label{11}
=\int e^{N\sum_{m>0} \frac{p_m}{m}\tr (M Z_2^\dag )^m} \det\left(1-  Z_1J_l Z_2\right)^{-Na}
d\mu(Z_1)d\mu(Z_2)d\nu(M)
\ee
$$
=\sum_{\lambda\cup\lambda\atop\ell(\lambda\cup\lambda)\le N}
(Na)_{\lambda\cup\lambda}(l)_{\lambda\cup\lambda}s_{\lambda\cup\lambda}(N\pb)N^{-2|\lambda|}
=\tau^{\rm DKP}(N\pb)
$$
where integrals (\ref{1})-(\ref{4}) are obtained from the graph $\Gamma_4$ where the each edge ends on different vertices and integrals (\ref{5}-\ref{11}) are obtained from the graph $\Gamma_5$ with one loop 
edge.

Similarly, one can write all realizations of the DKP tau function written according to these two graphs $\Gamma_4$ and $\Gamma_5$ in the so-called Miwa variables
\footnote{An example of a tau function written in Miva variables is the famous Kontsevich integral. It has a source matrix $\Lambda$, and the integral is a function of the eigenvalues of
$\Lambda$.}. We leave this as an exercise for the interested reader.

So, we can see that the set of models with the same partition function is a family,
whose elements differ in the choice of graph for given numbers $F, n, V$.
We can then permute the monodromies of faces and permute the monodromies of vertices. And we are free to choose where to place the matrix $M$, over which we additionally average.

\section{Discussion \label{discussion}} 

Now let us mention a set of problems related to the topic of this note.

\begin{itemize}
\item
The equality
\be\label{forDSE}
s_{\lambda\cup\lambda}(\pb_\infty)=
\frac{\prod_{i<j\le N}\left(x_i-x_j \right)^2\left((x_i-x_j)^2-1 \right)}{\prod_{i=1}^N x_i!(x_i-1)!},\quad x_i=\lambda_i-2i+1+2N,
\ee
makes series 
(\ref{borodin}), (\ref{*}) 
and other ones to be related to discrete symplectic ensembles studied in \cite{BorodinStrahov},\cite{Forrester-discrete}.

Other cases considered here result in modifications of discrete symplectic 
ensembles that have not been studied yet. 
It can be described as a one-dimensional log gas of pairs of particles ('dumbbells').
 For $\pb=\pb_\infty$ the sum in (\ref{*}) can be interpreted as the partition function for a system of hard core
particles on the lattice $\mathbb{Z}_{\le 0}$ coupled in dumbbells with mass 
$m= p-1$  and attracting according to two-dimensional gravity law.
The state with minimal energy is 
the dense package. The limiting shape for an infinite number  of particles and for a  special choice of external potential in the case of a finite temperature should be
obtained from the variational principle as it was shown in \cite{VershikKerov} 
for Coulomb lattice gas and as it was done, for instance, in \cite{OLeur-particles}).
It will be studied in a more detailed text.
Here, we only point out the appearance of
such discrete ensembles in the context of matrix models.

\item
It is interesting to understand the combinatorial meaning of sums in the right hand side
of (\ref{ZU-MM''}) similar to the meaning of Feynman graphs in the matrix models related
to unitary. orthogonal and symplectic ensembles and in the model (\ref{ZU-MM}) where
this meaning is well-studied. 
One can expect to obtain a variety of combinatorial equalities because the same expression will correspond to completely different Feynman diagrams due to ambiguity as described in Section 
\ref{equalities}.

\item Is it possible to apply the method of topological recursion \cite{Kazarian} to such matrix models?

\item  In addition to the ensemble of complex matrices on graphs, matrix models can be built from an ensemble of unitary matrices. They are united by the fact that both complex and unitary matrices form a group. I will hypothesize that multimatrix models with other ensembles, such as $\omega_1,\omega_2$ and $\omega_3$, can be built from graphs, and instead of Schur functions, the characters of the representations of these groups will be used.

\item We can conjecture that there is a generalization of equalities (\ref{ZU-MM})
and of (\ref{ZU-MM''}) to the other polynomials (for instance, Macdonald one's) and integrals
over quantum groups. At this point, it is clear how to construct matrix models on graphs with replacement of Schur functions by zonal polynomials (a note on this will be published separately, since the connection with tau functions and known integrable systems is not yet clear, while this paper is devoted to the connection with integrable systems).

\end{itemize}

\bigskip
\bigskip
\noindent
 \section{ Acknowledgements.}

With this article the author wants to honor the memory of the deceased teacher Vladimir Zakharov. He supported me greatly in my first steps in research work. I am happy that I use the results of his remarkable work in the theory of nonlinear wave equations and integrable systems.

This work is an output of a research project implemented as part of the Basic Research Program at the National Research University Higher School of Economics (HSE University).

\bigskip



\appendix

\section{Partitions. The Schur polynomials \cite{Mac} \label{Partitions}}
 
We recall that a nonincreasing set of nonnegative integers $\lambda_1\ge\cdots \ge \lambda_{k}\ge 0$,
we call partition $\lambda=(\lambda_1,\dots,\lambda_{l})$, and $\lambda_i$ are called parts of $\lambda$.
The sum of parts is called the weight $|\lambda|$ of $\lambda$. The number of nonzero parts of $\lambda$
is called the length of $\lambda$, it will be denoted $\ell(\lambda)$. See \cite{Mac} for details.
Partitions will be denoted by Greek letters: $\lambda,\mu,\dots$. The set of all partitions is denoted by
$\Pa$. The set of all partitions with odd parts is denoted $\OP$.
Partitions with distinct parts are called strict partitions, we prefer
letters $\alpha,\beta$ to denote them. The set of all strict partitions will be denoted by $\DP$.
The Frobenius coordinated $\alpha,\beta$ for partitions $(\alpha|\beta)=\lambda\in\Pa$ are of usenames
(let me recall that the coordinates $\alpha=(\alpha_1,\dots,\alpha_k)\in\DP$ consists of the lengths of arms counted
from the main diagonal of the Young diagram of $\lambda$ while
$\beta=(\beta_1,\dots,\beta_k)\in\DP$ consists of the lengths of legs counted
from the main diagonal of the Young diagram of $\lambda$, $k$ is the length of the main diagonal of $\lambda$,
see \cite{Mac} for details).

To define the Schur function $s_\lambda$, $\lambda\in\Pa$ at the first step we introduce the 
set of elementary Schur functions $s_{(m)}$ by
\be\label{CLit}
e^{\sum_{m>0}\frac 1m p_m x^m}=\sum_{m\ge 0} x^m s_{(m)}(\pb)
\ee
where the variables $\pb=(p_1,p_2,p_3,\dots)$ are called power sum variables. For 
$\lambda=(\lambda_1,\lambda_2,\dots)\in\Pa$
we define 
\be
s_\lambda(\pb)=\det\left[s_{(\lambda_i-i+j)}(\pb)  \right]_{i,j>0}
\ee
Let us introduce notations $\pb_\infty:=(1,0,0,\dots)$ and $\pb(a):=(a,a,a,\dots)$.
We have (see Ch. I in \cite{Mac})
\be\label{s(p(a))}
\frac{s_\lambda(\pb(a))}{s_\lambda(\pb_\infty)}=(a)_\lambda:=\prod_{(i,j)\in\lambda}(a-j+i)
\ee
The product ranges all nodes $(i,j)$ of the Young diagram $\lambda$ \cite{Mac}.
The product on the right-hand side is called the content product, since the number $j-i$
is called the content of the node with coordinates $(i,j)$.

The Schur functions play the role of the characters of the irreducable reperesentation 
labeled by $\lambda$ of the linear group.

If we put $p_m=p_m(X)=\ttr X^m$ where $X$ is an $N\times N$ matrix we write $s_\lambda(\pb(X))=s_\lambda(X)$.
If the rank of $X$ is equal to $l\le N$, for $\ell(\lambda)\le l$ we have
\be
s_\lambda(X)=\frac{\det\left(x_i^{\lambda_j-j+N}\right)_{i,j\le l}}
{\det \left(x_i^{-j+N}\right)_{i,j\le l}}=s_\lambda(\pb(X))
\ee
and $s_\lambda(X)=0$ if $\ell(\lambda)>l$.

Consider a matrix whose $l$ ($l\le N$) eigenvalues are equal to $1$ and $N-l$ eigenvalues are $0$.
Then, according to (\ref{s(p(a))}) we have
\be\label{s(J_l)}
s_\lambda(J_l)=(l)_\lambda s_\lambda(\pb_\infty)
\ee
In particular $s_\lambda(I_N)=(N)_\lambda s_\lambda(\pb_\infty)$.

The following relation is known as the character map relation  
\be\label{charmap}
s_\lambda(X)=s_\lambda(\pb_\infty)\sum_{\mu\atop |\mu|=|\lambda|}\varphi_\lambda(\mu)
\prod_{i=1}^{\ell(\mu)}
\tr\left( X^{\mu_i} \right),\quad \pb_\infty:=(1,0,0,\dots),
\ee
where $\varphi_\lambda(\mu)$ are rational numbers which are expressed in terms of 
the characters of the symmetric group $S_d,\,d=|\lambda|$, see \cite{Mac}. It follows from 
(\ref{charmap}) that $s_\lambda(X)$ is a polynomial of degree $d=|\mu|=|\lambda|$ in entries 
of $X$. Therefore, $s_\lambda(X)$ is a polynomial of each entry of matrix $X$ and one can integrate it over entries, as it was written in (\ref{Schur-mean-SE}),(\ref{Schur-mean-OE}),
 (\ref{Schur-mean-qGE}),(\ref{quatr-measure}), (\ref{1MM}), (\ref{I}), (\ref{complexMat-measure}).

 \section{Matrix ensembles\label{ME}}
 
 \paragraph{Symplectic matrices.}
The Haar measure on $\mathbb{S}p_{2N}$ is 
\be
d_*M=
\frac{2^{(N)^2}}{\pi^N}\prod_{i<j}^N \left(\cos\theta_i-\cos\theta_j  \right)^2
\prod_{i=1}^{N}\sin^2\theta_i d\theta_i ,
\ee 
 where
$d\theta_i ,\quad 0\le \theta_i <\cdots<\theta_N \le \pi$. The prefactors are chosen to provide 
$\int_{\mathbb{S}p_{2N}} d_*M=1$.

 \paragraph{Orthogonal matrices.}
The Haar measure on $\mathbb{O}_N$ is 
\be
d_*M=\begin{cases}
\frac{2^{(n-1)^2}}{\pi^n}\prod_{i<j}^n \left(\cos\theta_i-\cos\theta_j  \right)^2\prod_{i=1}^{n}d\theta_i,\quad N=2n
      \\
\frac{2^{n^2}}{\pi^n}\prod_{i<j}^n \left(\cos\theta_i-\cos\theta_j \right)^2 \prod_{i=1}^{n}
 \sin^2\frac{\theta_i}{2}  d\theta_i ,\quad N=2n+1
     \end{cases}
\ee 
 where
$d\theta_i ,\quad 0\le \theta_i <\cdots<\theta_n \le \pi$. The prefactors are chosen to provide 
$\int_{\mathbb{O}_N} d_*M=1$.

\paragraph{Complex matrices.}
The measure on the space of complex matrices is defined as 
\be\label{complexMat-measure}
d\mu(Z)=c_N \prod_{a,b=1}^N d\Re Z_{ab}d\Im Z_{ab}\text{e}^{-N|Z_{ab}|^2}
\ee

Consider integrals over $N\times N$ complex matrices $Z_1,\dots,Z_n$ where the measure is defined as
\be\label{CGEns-measure}
d\mu(Z_1,\dots,Z_n)= c_N^n
\prod_{i=1}^n\prod_{a,b=1}^N d\Re (Z_i)_{ab}d\Im (Z_i)_{ab}\text{e}^{-N|(Z_i)_{ab}|^2}
\ee
where the integration domain is $\mathbb{C}^{N^2}\times \cdots \times\mathbb{C}^{N^2}$ and where $c_N^n$
is the normalization
constant defined via $\int d \Omega(Z_1,\dots,Z_n)=1$.

\end{document}